# FROM 3D POINT CLOUDS TO SEMANTIC OBJECTS
*An Ontology-Based Detection Approach*

Helmi Ben Hmida[1,2], Christophe Cruz[2], Frank Boochs[1], Christophe Nicolle[2]

[1] *Institut i3mainz, am Fachbereich 1 - Geoinformatik und Vermessung*
*Fachhochschule Mainz, Lucy-Hillebrand-Str. 2, 55128 Mainz*
*{helmi.benhmida, boochs}@geoinform.fh-mainz.de*

[2] *Laboratoire Le2i, UMR-5158 CNRS,*
*Dep. Informatique IUT Dijon, 7, Boulevard Docteur Petitjean*
*BP 17867, 21078 Dijon CEDEX, France*
*{christophe.cruz, cnicolle}@u-bourgogne.fr*

Abstract: This paper presents a knowledge-based detection of objects approach using the OWL ontology language, the Semantic Web Rule Language, and 3D processing built-ins aiming at combining geometrical analysis of 3D point clouds and specialist's knowledge. This combination allows the detection and the annotation of objects contained in point clouds. The context of the study is the detection of railway objects such as signals, technical cupboards, electric poles, etc. Thus, the resulting enriched and populated ontology, that contains the annotations of objects in the point clouds, is used to feed a GIS systems or an IFC file for architecture purposes.

## 1 INTRODUCTION

With every new scanner model on the market, phase-shift scanners in particular, the instruments become faster, more accurate and can scan objects at longer distances. Laser scanners tests which have been carried out at the i3mainz institute for eight years now, prove this observation (Boehler, Bordas Vicent, & Marbs, 2003). Surveying with 3D scanners is spreading all domains during the past decade. Terrestrial laser scanners have been established as a workhorse for topographic and building survey from the archaeology (Balzani, Santopuoli, Grieco, & Zaltron, 2004) to the architecture (Hammoudi, 2009). Actually, such technology presents a powerful tool for many applications and has partially replaced traditional surveying methods since it can speed up field work significantly. Actually, this powerful method allows the creation of 3D point clouds from objects or landscapes. However, the huge amount of data generated during the process proved to be costly in post-processing. The field time is very height since in most cases; processing techniques are still mainly affected by manual interaction of the user. Typical operations consists to clean point clouds, to delete unnecessary areas, to navigate in an often huge and complicated 3D structure, to select set of points, to extract and model geometries and objects. At the same time it would be much more effective, to process the data automatically, which has already been recorded in a very fast and effective way.

As object reconstruction is an important task for many applications, considerable effort has already been invested to reduce the impact of time consuming, manual activities and to substitute them by numerical algorithms. Actually, the automatic processing of 3D point clouds can be very fast and efficient, but often relies on significant interaction of the user for controlling algorithms and verifying the results. Alternatively, the manual processing is intelligent and very precise since a human person uses its own knowledge for detecting and identifying objects in point clouds, but it is very time-consuming and consequently inefficient and expensive. In this context, we aim at inserting business knowledge in automatic detection and reconstruction algorithms in order to make the point cloud processing more efficient and reliable.

Consequently, the WiDOP project (knowledge based detection of objects in point clouds) aims at making a step forward. The goal is to develop efficient and intelligent methods for an automated processing of terrestrial laser scanner data. In contrast to existing approaches, the project consists in using prior knowledge about the context and the objects. This knowledge is extracted from databases, CAD plans, Geographic Information Systems (GIS), or domain experts. Therefore, this knowledge is the basis for a selective knowledge-oriented detection.

The project WiDOP is Funded by the German government. However, the partners are the German railway company (Deutsche Bahn), the Fraport company (Frankfurt Airport manager), and the Metronom company which is specialized in 3D point cloud processing. The Fraport company main concerns are building and furniture management of the airport. The furniture's position relative to the security gates is constantly moving. In addition, updates are done on buildings such as new walls, destruction of walls, new holes in a wall, new windows, etc. This could be undertaken by the director of a new shop or by the technical employers in order to reorganize storerooms for instance. As a matter of fact, it is very difficult to keep up to date the plans of the airport. The Deutsche Bahn main concerns are the management of the railway furniture. Actually, the environment of the railway is constantly changing. And the cost of keeping these plans up to date is increasing. The present-time solution adopted by the Deutsche Bahn (DB) consists on fixing a 3D terrestrial laser scanner on the train and to survey the surrounding landscape (Railway, signals and green trees on the borders). Metronom automation is a DB subcontractor specialized in 3D data processing. This partner takes the survey point clouds as input and detects the different existent elements manually helped with some 3D process like spike detection. The main objective of Deutsch Bahn project consists in detecting automatically the objects in the 3D point clouds in order to feed the position and the semantic definition of objects into a GIS system.

The following paper is structured into section 2 which gives an overview of actual existing strategies for reconstruction processes, section 3 explains the general adopted architecture and the related ontology structure, section 4 describe the domain knowledge modelling, section 5 highlight the annotation process, section 6 gives first results for a real example and section 7 concludes and shows next planned steps.

## 2 BACKGROUND

This section is composed of two parts. This first part deals with the detection strategies described in the literature for geometric modelling and object recognition. The second part presents the knowledge modelling which of value for our strategy.

### 2.1 Detection strategies

Today, scene model creation process is largely a manual procedure, which is time-consuming and subjective. While there is a clear need for automated, or even semi-automated methods to ease the creation of as-built scene, research on the subject is still in the very early stages. This survey shows that many of the existing methods for geometric modelling and object recognition can be important for the process automation. Within the literature, three main strategies are described where the first one is based on human interaction with provided software's for point clouds classifications and annotations. While the second strategy relies more on the automatic data processing without any human interaction by using different segmentation techniques for features extraction. Finally, new techniques present an improvement compared with the cited ones by integrating semantic networks to guide the reconstruction process.

#### 2.1.1 Manual Supported strategy

Actually, tools used for 3D reconstruction of objects are still largely relying on human interaction. Here the user might be supported in his construction activity, but object interpretation, selection and extraction of measurements has to be done manually. That's why this processing is the most time consuming way to come from a data set to extracted objects (Leica Cyclone: 3D Point Cloud Processing Software).

#### 2.1.2 Semi-automatic and automatic strategy

These methods present a real optimization within the process compared of the manual ones. Within the current section, we will not expose the problematic from the automatism point of view, but these methods are based on two main parts, geometry extraction and annotation.

Basically, geometry extraction presents the process of constructing a simplified representation of a 3D shape such as a Signal or an Electric born like in our case. The representation of geometric shapes has been studied extensively, (Campbell & Flynn,

2001). In our context, shape representations can be classified along three independent dimensions: parametric versus nonparametric, global versus local, and explicit versus implicit. Parametric representations describe a shape using a model with a small number of parameters. Non-parametric representations do not have any parameters. A cylinder in this case can be represented using a triangle mesh. In this area, the output can be a surface based like a boundary based representation, or a volumetric representation, (Curless & Levoy, 1996) like the case of CSG representation. The model can be represented by combining some fixed primitives with Boolean operators (union, difference, and intersection). Global representations describe the shape of an entire object, while local representations may characterize only a portion. The dimension of explicit versus implicit representations is perhaps the most significant axis for distinguishing shape representations. Explicit representations directly encode the shape of an object (e.g., a triangle mesh), while implicit representations indirectly encode object shape using an intermediate representation. Explicit representations are well suited for modelling 3D objects, whereas implicit representations are most often used for 3D object recognition and classification. In this field, there has been some work on detecting and modelling more complex structures. These methods often include some aspects of object recognition or depend on a prior knowledge of object class (Faber & Fisher, 2002), and a genetic algorithm to fit parametric models of doors point clouds for instance. Pu and Vosselman used a triangulation-based method to detect the boundaries of sparse regions within a building façade and then fit rectangles to the resulting regions (Pu & Vosselman). Böhm et al. use density-based edge detection to find vertical and horizontal lines in the depth map of a building façade.

Once geometric elements are detected and stored via a specific presentation, the second core of the object detection and scene reconstruction is object recognition, In fact, it presents the process of labelling a set of data points or geometric primitives extracted from the data with a named object or object class. Whereas the geometry modelling task would find a set of points to be a vertical Bounding Box, the recognition task would label that Box as a Signal. Object recognition algorithms may label object instances of an exact shape, or they may recognize classes of objects. Research on recognition of specific building components is still in its early stages. Methods in this category are typically shape-based ones. They aim at segmenting a scene into planar regions, for example, and then use features derived from the segments to recognize objects. This approach was carried out by Rusu et al. by using heuristics to detect walls, floors, ceilings, and cabinets in a kitchen environment, (Rusu, 2008). A similar approach was proposed by Pu and Vosselman to model building façades, (Pu S. a., 2009). One of the challenges of recognition in the building context is that many of the objects to be recognized are very similar to objects of little relevance. Some researchers have proposed qualifying the spatial relationships between objects or geometric primitives to reduce the ambiguity of recognition results. Such approaches generate semantic labels of geometric primitives, and test the validities of these labels with a spatial relationship knowledge base. Usually, such a knowledge model is represented by a semantic network, (Nuchter, 2008). For instance, a semantic net may specify the relationships between entities such as "floors are orthogonal to walls and doors, and parallel with ceilings". Such validity checking approaches provide ways to integrate domain knowledge into the object recognition process. Another approach for recognition is to first detect objects that are easily recognizable, and then use the context of these initial detections to facilitate recognition of more challenging structures. For example, Pu and Vosselman use characteristic features, such as size, orientation, and relationships to other prominent objects, to detect walls and roofs (Pu S. a., 2009). Then, a second stage detects windows within each of the detected walls.

One strategy for reducing the search space of object recognition algorithms is to utilize knowledge about a specific facility, such as a CAD model or floor plan of the original design. For instance, Yue et al. overlay a design model of a facility with the as-built point cloud to guide the process of identifying which data points belong to specific objects and to detect differences between the as-built and as-designed condition (Yue, 2006). In such cases, object recognition problem is simplified to be a matching problem between the scene model entities and the data points. Another similar approach is presented in (Osche, 2008).

From the above mentioned works, we can deduce that the problematic of 3D object detections and scene reconstructions including standard algorithm and semantic networks can produce first results. Moreover such strategies suffer from the lack of flexibility, efficiency and are in general hard coded. Thus, the context and the algorithm which are part of knowledge that are required to be used in recognition process have to be modelled.

## 2.2 Knowledge modelling

In recent years, formal ontology has been suggested as a solution to the problem of 3D objects reconstruction from 3D point clouds (Cruz, Marzani, & Boochs, 2007). In this area, ontology structure was defined as a formal representation of knowledge by a set of concepts within a domain, and the relationships between those concepts. It is used to reason about the entities within that domain, and may be used to describe the domain. Conventionally, ontology presents a "formal, explicit specification of a shared conceptualization" (Gruber, 2005). Ontology provides a shared vocabulary, which can be used to model a domain. Well-made ontology own a number of positive aspects like the ability to define a precise vocabulary of terms, the ability to inherit and extends exiting ones, the ability to declare relations ship between defined concepts and finally the ability to infer new relationship by reasoning on existing ones. Through technologies known as Semantic Web, most precisely the Ontology Web Language (OWL) (McGuinness & Harmelen, 2004), researcher are able to share and extends knowledge through the scientific community. The basic strength of formal ontology is their ability to reason in a logical way based on Description Logics DL. The last one presents a form of logic to reason on objects. Lots of reasoners exist nowadays like Pellet (Sirin, Parsia, Grau, Kalyanpur, & Katz, 2007), (Tsarkov & Horrocks, 2006) and KAON (U. Hustadt, 2010). Despite the richness of OWL's set of relational properties, the axioms does not cover the full range of expressive possibilities for object relationships that we might find, since it is useful to declare relationship in term of conditions or even rules. These rules are used through different rules languages to enhance the knowledge possess in an ontology. In the last few years, lots of rules languages have been emerged. Some of the evolved languages are related to the semantic web rule language (SWRL) and advanced Jena rules (Carroll, Dickinson, Dollin, Reynolds, Seaborne, & Wilkinson, 2004). SWRL is a proposal as a Semantic Web rules language, combining sublanguages of the OWL Web Ontology Language with the Rule Markup Language (Horrocks, Patel-Schneider, Boley, Tabet, Grosof, & Dean, 2004). A famous example rules would be to assert that the combination of the `hasParent` and `hasBrother` properties implies the `hasUncle` one. This rule could be written as:

```
hasParent(?x1,?x2)^hasBrother(?x2,?x3)→
hasUncle(?x1,?x3)
```

Where x1, x2 and x3 represent the individuals of the class `Person` defined in the ontology and `hasParent`, `hasBrother`, and `hasUncle` presents data property or relations between individuals in the same cited class. As seen in the above example, rules are divided in two parts, antecedent and consequent separated by the symbol "→". If all the statement in the antecedent clause is determined to be true, then all the statement in the consequent clause is applied. In this way, new properties like `hasUncle` in our example can be assigned to individuals in the ontology. In addition, SWRL language specifies also a library for mathematical built-ins functions which can be applied to individuals. It includes numerical comparison, simple arithmetic and string manipulation.

In this project, domain ontologies are used to define the concepts, and the necessary and sufficient conditions that define the concepts. These conditions are of value, because they are used to populate new concepts. For instance, the concept "Horizontal_BoudinBox" can be specialized into "Wall" if it contains a "Window". Consequently, the concept "Wall" will be populated with all "Horizontal_BoudinBox" if they are linked to a "Window" or "OpeningElement" object (Vanlande, 2008). In addition, the rules are used to compute more complex results such as the topological relationships between objects. For instance, the intersection of two objects is used to determine if a part of an object is inside of another object. The ontology is than enriched with this new relationship. The topological relation built-ins are not defined in the SWRL language. Consequently, the language was extended.

## 3 APPROACH OVERVIEW

This paper presents a knowledge based detection approach using the OWL ontology language, the Semantic Web Rule Language, and 3D processing built-ins aiming at combining geometrical analysis of 3D point clouds and specialist's knowledge. This combination allows the detection and the annotation of objects contained in point clouds. The field of the Deutsch Bahn railway scene is treated for object detection. The objective of the system consists in creating, from a set of point cloud files, from an ontology that contains knowledge about the DB railway objects, and from the knowledge about 3D processing algorithms, an automatic process that produces as output a set of tagged elements contained in the point clouds. The following picture shows elements that can be found those cloud points.

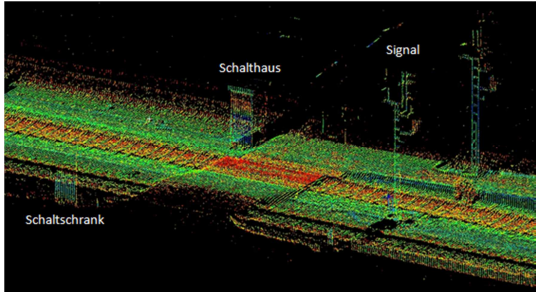

Figure 1: Deutsche Bahn scene point clouds

The process enriches and populates the ontology with individuals and relationships between these new individuals. In order to graphically represent these objects, a VRML file (VRML Virtual Reality Modeling Language, 1995) is generated. In addition, the colour of objects in the VRML file represents its semantic definition. The resulting ontology contains enough knowledge to feed a GIS system, and to generate IFC file (IFC Model, 2008) for CAD software, but this is out of the scope the paper.

The processing steps can be detailed within the schema of Figure 2, where three main steps aim at detecting and identifying objects.
    (3) From 3D point clouds to geometric elements.
    (4) From geometry to topologic relations.
    (5) From geometric and/or topologic relations to semantic elements annotation.

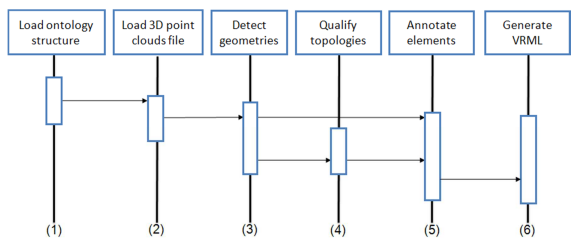

Figure 2. Sequence of the object detection application

As intermediate steps, the different geometries within a specific 3D point clouds are detected and stored within the ontology structure. Once done, the existent topological relations between the detected geometries are qualified and then stored within the same knowledge base. Finally, detected geometries are annotated semantically, based on existing knowledge's related to the geometric characteristics and topologic relations. The input ontology contains knowledge about the DB railway objects and knowledge about 3D processing algorithms. Consequently, the knowledge base is divided into two layers, the layer of DB object description and the layer of the algorithmic description.

The object description into the ontology is classified in three sub-layers: geometric, topologic and scene knowledge. The sub-layer of scene knowledge is composed by three main classes which are the Scene, the domain concepts and the characteristics. The second one can be considered as the main class since it presents the target semantic elements and contains all relevant object elements which might be found within that scene. In case of Deutsche Bahn scene, this might comprise a list such as: {Signals, Mast, Schalanlage, etc.}. Besides, the importance of the other classes cannot be ignored. They are used to fix either the main scene within its point clouds file and its size through attributes related to the scene class, or even to characterize detected element with different semantic and geometric characteristics.

The sub-layer of the geometrical knowledge formulates the basic geometrical elements used within the prototype. Actually, the annotation elements step processes bounding boxes. Other geometries especially lines and planes are more used to characterize domain concepts elements by a list of geometries. This information is used to create useful descriptions that facilitate the object detection process. For instance, an electric pole (Mast) is composed by vertical lines connected to horizontal ones.

The sub-layer of the topologic knowledge represents topological relationships between scene elements. For instance, a topological relation between a distant signal and a main one can be defined, as both have to be distant of 1 Km. The qualification of topologic relations into the semantic framework is done by means of topological Built-Ins called "3DSWRL_Topologic_Built-Ins". The connections of domain concept to the geometries and the characteristic classes or others are carried out through the object properties. Further, the object properties are also used to link an object to others by a topologic relation. In general there are a set of object properties in the ontology which have their specialized properties for the specialized activities.

The 3D processing algorithmic layer contains all relevant aspects related to the 3D processing algorithms. It´s integration into the semantic framework is done by means of special Built-Ins called "Processing Built-Ins". They manage the interaction between above mentioned layers. In addition, it contains algorithm definitions, properties, and geometries related to the each defined algorithms. An importance achievement is the detection and the identification of objects which has linear structure such as signal, indicator column, and electric pole, etc., through utilizing their geometric properties. Since the information in point cloud data sometimes is unclear and insufficient, thus the various methods to RANSAC, (Tarsha-Kurdi, Landes, & Grussenmeyer, 2007) are combined and upgraded. This combination is able to robustly detect the best fitting lines in 3D point clouds for example.

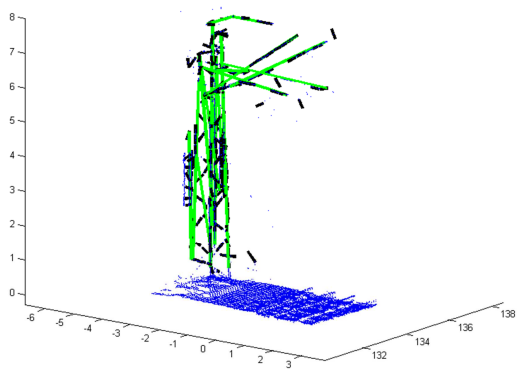

Figure 3: Mast detection

Figure 3 contains the Mast object constructed by linear elements, ambiguously represented in point cloud as blue points. Green lines are results of possible fitting lines and clearly show the shape of the object that is defined in the ontology. The object generated from this part is a bounding box that includes all inside geometries of the object and a concept label. The description of this layer is out of the scope of the paper, since we focus here on the knowledge modelling part.

### 3.1 Ontology schema

This section discusses the different aspects related to the ontology structure installed behind the WiDOP prototype to respond to the different cited purposes. The domain ontology presents the core of WiDOP project and provides a knowledge base to the created application. The global schema of the modelled ontology structure offers a suitable framework to characterize the different Deutsche Bahn elements from the 3D processing point of view. The created ontology is used basically for two purposes:

- To guide the processing algorithm sequence creation based on the target object characteristics.
- To facilitate the semantic annotation of the different detected objects inside the target scene.

The current ontology is divided onto three main parts: the Deutsche Bahn concepts, the algorithm concepts and the geometry concepts. However, they will be used with others to facilitate the object detection based on SWRL and the automatic annotation of Bounding Box geometry based on inference engine tools. At this level, no real interaction between human and the knowledge base in taken in consideration, since the 3D detection process algorithm and parameters are alimented directly from the knowledge base and then interpreted by the SWRL rules and Description Logics tools. The ontology is managed through different components of Description Logics. There are five main classes within other data and objects properties able to characterize the scene in question.

- Algorithm
- Geometry
- DomainConcept
- Characteristics
- Scene

The class DomainConcept can be considered the main class in this ontology as it is the class where the different elements within a 3D scene are defined. It was designed after the DB scene observation. It contains all kinds of elements, which have to be detected and is divided in two general classes, one for the Furniture and one for the Facility Element. However, the importance of other classes cannot be ignored. They are used to either describe the domain concept geometry and characteristics or to define the 3D processing algorithms within the target geometry. The subclasses of the Algorithm class represent the different developed algorithms. They are related to several properties which they are able to detect. These properties (Geometric and semantic) are shared with the DomainConcept and the Geometry classes. By this way, a created sequence of algorithms can detect all the characteristics of an element while the Geometry class represents every kind of geometry, which can be detected in the point cloud scene.

The connection between the basic mentioned classes is carried out through object and data properties. There exist object properties for each mentioned activities. Besides, the object properties are also used to relate an object to other objects via topological relations. In general, there are five general object properties in the ontology which have their specialized properties for the specialized activities. They are

- hasTopologicRelation
- IsDeseignedFor
- hasGeometry
- hasCharacteristics,

Figure 4 demonstrates the general layout schema of the application.

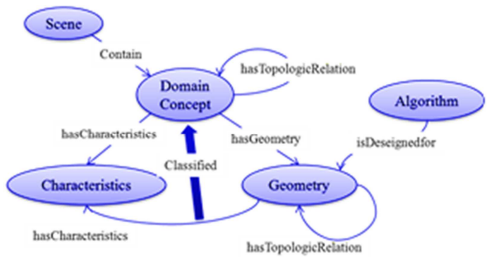

Figure 4. Ontology general schema overview

## 4 DOMAIN KNOWLEDGE MODELLING

The created knowledge base related to the Deutsche Bahn scene has been inspired next to our discussion with the domain expert and next to our study based on the official Web site for the German rail way specification "http://stellwerke.de". Once done, new knowledge related to the different element´s geometry and metrological relations between them is synthetized. An overview of the targeted elements, the most useful and discriminant characteristics to detect it and their inter-relationship is presented.

Basically, a railway signal is one of the most important elements within the Deutsche Bahn scene where we find main signals and secondary ones. The main signals are classified onto primary signal and the distant ones. In fact, the primary signal is a railway signal. It indicates whether the subsequent section of track may be driven on. A primary signal is usually announced through a distant signal. The last one indicates which image signal to be expected will be associated to the main signal in a distance of 1 km. Big variety of secondary signals exists like the Vorsignalbake, the Haltepunkt and others. From the other side, the other discriminant elements within the same scene are the Masts presenting electricity born for the energy alimentation. Usually, masts are distant of 50 m from each other's. Finally, the Schaltanlage elements presents small electric born connected to the ground.

### 4.1 Geometric characteristics

Geometric characteristics can present a discriminant feature able to improve the automatic annotation process. For this reason, we opt to study the different geometric features related to the cited elements, then, use only the discriminant one as a basic features for a given object. As a first step, bounding boxes are used in order to focus on topological relationships. The following table groups the object characteristics together regarding the properties of a bounding box (Figure 5). This table is extended with algorithm characteristics, but it is not presented here.

| | | | BB Height | BB Length | BB Width |
|---|---|---|---|---|---|
| Signals | Primary signal | Main Signal | Between 4 and 6m | | |
| | | Distant Signal | Between 4 and 6m | | |
| | Secondary signal | Vorsignalbake | between 1,5 and 2.5m | | Close to 0 |
| | | Breakpoint_table | between 1 and 2m | between 1 and 1,5m | Close to 0 |
| | | Chess_board | between 1 and 1,5m | Close to 0 | Close to 0 |
| Mast | | BigMast | More than 6m | | |
| | | NormalMast | Between 5 and 6 | Close to 0 | |
| Schaltanlage | | Schalthause | Less than 1m | | |
| | | SchaltSchrank | Less than 0,5m | | |

Figure 5: Geometric characteristics overview

### 4.2 Topologic characteristics

While exploring the railway domain, lots of standard rules are imposed; such rules are used to help the driver and to ensure the passengers security. From our point of view, these are helpful also to verify and to guide the annotation process. For instance, the distance between the distant signal and the main one corresponds to the stopping distance that the trains require. The stopping distance shall be set on specific route and is in the main lines often 1000 m or in rare case 700 m. Add to that, three to five Vorsignalbake are distant around 75m while then the last one is distant of 100m to the distant signal, Figure 6.

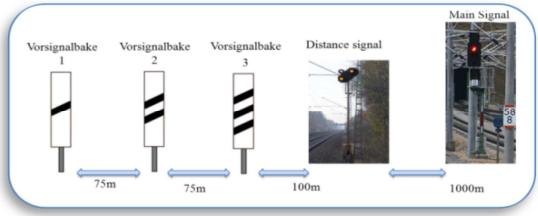

Figure 6: Topologic rules

Same thing related to the mast elements, since usually, they are distant of 50 m from each other's in the normal cases. In addition, an important other information able to discriminate the masts from the signal is related to the small electricity box always connected to the signal bottom.

The next section introduces an overview of the approach undertaken in the WiDOP project to detect and annotate semantically the different Deutsch Bahn objects.

# 5 SEMANTIC ANNOTATION PROCESS

The process that consists to qualify the topological relationships between geometries is based on SWRL rules.

## 5.1 Point cloud to geometry

The first step aims at the geometric elements detection. Thus, Semantic Web Rule Language within extended built-ins for complex 3D processing are used in order to detect geometry (e.g. Table 1). Once done, the detected elements are used to populate the ontology.

The "3Dswrlb:VerticalElementDetection" built-ins aims at the detection of vertical elements.

The prototype of the designed Built-in is:

```
3D_swrlb_Processing:
VerticalElementDetection(?Vert, ?Dir)
```

where the first parameter presents the target object class, and the last one presents the point clouds directory defined within the created scene. At the moment, the detection process will result bounding boxes, representing a rough position and orientation of the detected object. Table 1 show the mapping between the 3D processing built-ins, which are computer and translated to predicate, and the corresponding class.

Table 1. 3D processing built-Ins mapping process

| Processing Built-Ins | Correspondent class |
|---|---|
| 3D_swrlb_Processing: VerticalElementDetection(?Vert,?Dir) | Vertical_BoundingBox(?x) |
| 3D_swrlb_Processing: HorizentalElementDetection(?Vert,?Dir) | Horizental_BoundingBox(?y) |

## 5.2 Geometries to topology

Once geometries are detected, the second step, aims at verifying certain topology properties between detected geometries. Thus, 3D_Topologic built-ins have been added in order to extend the SWRL language. Topological rules are used to define constrains between different elements. After parsing the topologic built-ins and its execution, the result is used to enrich the ontology with relationships between individuals that verify the rules. Similarly to the 3D processing built-ins, our engine translates the rules with topological built-ins to standard rules, Table 2.

Table 2. Example of topologic built-ins

| Processing Built-Ins | Correspondent object property |
|---|---|
| 3D_swrlb_Topology:Upper (?x, ?y) | Upper(?x,?y) |
| 3D_swrlb_Topology:Intersect(?x, ?y) | Intersect (?x,?y) |

## 5.3 Geometry to semantic

After the geometry and the topological relation detection, rules aim at qualifying and annotating the different detected geometries. These rules control and manage the annotation process. The following example shows how a rule specifies the class of a VerticalElement which is of type Mast regarding its altitude. The altitude is highly relevant only for this element.

```
3DProcessing_swrlb:VerticalElementDetec
tion(?Vert, ?dir) ^ altitude (?x, ?alt)
^swrlb:moreThan (?alt, 6) → Mast
(?Vert)
```

## 5.4 Topology to semantic

In other cases, geometric knowledge is not sufficient for the previous process. The topologic relationships between detected geometries are helpful to manage the annotation process. The following example shows how semantic information about existing objects is used conjunctly with topological relationships in order to define the class of another object.

```
Mast (?vert1) ^ VerticalBB (?Vert2) ^
hasDistanceFrom (?vert1,?vert2, 50) →
Mast(?vert2)
```

## 6 CASE STUDY

For the demonstration of our system, 500 m from the scanned point clouds related to Deutsch Bahn scene in the city of Nürnberg was extracted. It contains a variety of the target objects. The whole scene has been scanned using a terrestrial laser scanner fixed within a train, resulting in a large point cloud representing the surfaces of the scene objects.

Different rules are processed. First, all vertical elements will be searched in the area of interest, and then topological relations between detected geometries are qualified. Subsequently further annotation may be relayed on aspects expressing facts to orientation or size of elements, which may be sufficient to finalize a decision upon the semantic of an object or on fact expressing topologic relationship or both of them.

This second step within our approach aims to identify existing topologies between the detected geometries. To do, useful topologies for geometry annotation are tested. Topologic Built-Ins like `isConnected`, `touch`, `Perpendicular`, `isDistantfrom` are created. As result, relations found between geometric elements are propagated into the ontology, serving as an improved knowledge base for further processing and decision steps.

The last step consists in annotating the different geometries. Vertical elements of certain characteristics can be annotated directly.

In more sophisticated cases, the combination of semantic information and topologic ones can deduce more robust results by minimizing the false acceptation rate. Finally, based on a list of SWRL rules, most of detected geometries are annotated. In this example, among 67 elements are classified as Masts, 21 SchaltAnlage, 34 basic signals and finally, 155 secondary signals.

The created platform offer the opportunity to materialize the annotation process by the generation and the visualization on a VRML structure alimented from the knowledge base. It ensures an interactive visualization of the resulted annotation elements beginning from the initial state, to a set of intermediate states coming finally to an ending state, Figure 7 where the set of rules are totally executed.

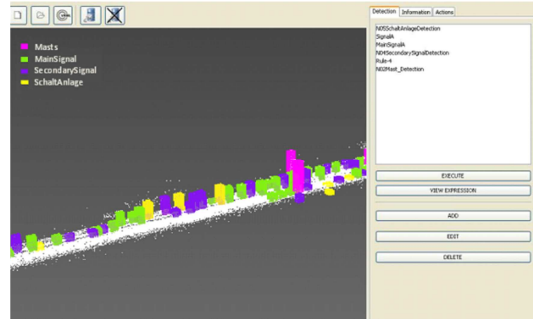

Figure 7. Detected and annotated elements visaliazation within VRML language

## 7 CONCLUSIONS

We have proposed a new solution to perform the detection of objects from technical survey within the laser scanner technology. The solution performs the detection of objects in 3D point clouds by using available knowledge about a specific domain (DB). This prior knowledge modelled within ontology SWRL rules are used to control the 3D processing execution, the topologic qualification and finally to annotate the detected elements in order to enrich the ontology and to drive the detection of new objects.

Future work will include the integration of new knowledge's that can intervene within the annotation process like the number of detected lines within each bounding box and the update of the general platform architecture, by ensure more communication between the scene knowledge within the 3D processing.

## ACKNOWLEDGEMENTS

This paper presents work performed in the framework of research project funded by the German ministry of research and education under contract No. 1758X09. The authors cordially thank for this funding. Special thinks also for Hung Truong, Yoann Rous and Romain Tribotté for their contribution for the Java prototype and C++ code development respectively.